# Precessing ball solitons in kinetics of a spin-flop phase transition

**V.V.Nietz**

*Frank and Shapiro Laboratory of Neutron Physics, Joint Institute for Nuclear Research, Dubna, Moscow Region 141980, Russia*

## Abstract

The fundamentals of precessing ball solitons (PBS), arising as a result of the energy fluctuations during spin-flop phase transition induced by a magnetic field in antiferromagnets with uniaxial anisotropy, are presented. The PBS conditions exist within a wide range of amplitudes and energies, including negative energies relative to an initial condition. For each value of the magnetic field, there exists a precession frequency for which a curve of PBS energy passes through a zero value (bifurcation point), and hence, in the vicinity of this point the PBS originate with the highest probability. The characteristics of PBS, including the time dependences of configuration, energy, and precession frequency, are considered. As a result of dissipation, the PBS transform into the macroscopic domains of a new phase.

*PACS*: 75.25; 75.50.E; 29.30.H

*Keywords*: Antiferromagnet, Spin-flop transition, Magnetic field, Precession, Precessing ball soliton, Evolution of soliton, Kinetics of the phase transition.

Tel: +7-496 21-65-552
Fax: +7-496-21-65-882
*E-mail address*: nietz@nf.jinr.ru

## Introduction

Magnetic solitons with spherical symmetry, which can arise in crystals with magnetic ordering during the phase transitions induced by a magnetic field, were considered in some papers [1-8]. The cases when it is possible to confine oneself to uniaxial symmetry [4-8] are of particular interest. For such a crystal, in addition to the amplitude (with corresponding configuration) and pulse, the solitons have the third parameter: the frequency of precession. In these cases, each value of an external field relates to a continuous spectrum of solitons by frequency and corresponding energy. Near boundaries of metastability, for each value of a field, the spectrum of precessing ball solitons (PBS) has the bifurcation point where the probability of PBS spontaneous origin is increasing abnormally. The frequency dependencies of energy and PBS configuration, as well as the process of PBS spontaneous origin at spin-flop transition in antiferromagnets have been discussed in [8]. However, the analysis of time evolution connected with dissipation of energy in PBS is not correct in this article.

In the present article, it is shown that dissipation of energy is accompanied not only by the change of configuration of solitons but also by the change of the precession frequency. Taking into account this time dependency, we can carry out a more comprehensive analysis of the quantities of PBS and to consider the whole process of transformation of PBS into macroscopic domains of a new phase.

In the first chapter, we represent a more correct analysis of equations and expressions for PBS than in [7, 8]. In the second one, the character of the time change of PBS that connected with dissipation is shown. Then, in the third chapter, we analyze the so-called "equilibrium PBS". In the fourth chapter, the PBS in overcritical range of a field, i.e. outside the metastability region, are considered. In the fifth chapter, the influence of the movement on the form of PBS is discussed. Finally, in the sixth chapter, we estimate the influence of the demagnetizing fields.

## 1. Equations for precessing ball solitons (PBS)

To analyze magnetic solitons in an antiferromagnet with uniaxial anisotropy, we used the following expression for the macroscopic energy (as in [8]):

$$W = 2M_0\int\left\{-\frac{A}{2}|\mathbf{l}|^2+\frac{B}{2}|\mathbf{m}|^2+\frac{C}{4}(\mathbf{ml})^2+\frac{K_1}{2}\left(|m_\perp|^2+|l_\perp|^2\right)-\frac{K_2}{4}\left(|m_\perp|^2+|l_\perp|^2\right)^2-m_zH_z+\right.$$
$$\left.+\frac{\alpha_{xy}}{2}\left[\left(\frac{\partial\mathbf{m}}{\partial X}\right)^2+\left(\frac{\partial\mathbf{m}}{\partial Y}\right)^2+\left(\frac{\partial\mathbf{l}}{\partial X}\right)^2+\left(\frac{\partial\mathbf{l}}{\partial Y}\right)^2\right]+\frac{\alpha_z}{2}\left[\left(\frac{\partial\mathbf{m}}{\partial Z}\right)^2+\left(\frac{\partial\mathbf{l}}{\partial Z}\right)^2\right]\right\}dXdYdZ \quad (1)$$

($K_1 > 0, K_2 > 0$; magnetic field $H$ is directed along the anisotropy axis $Z$). This equation can be reduced to the following view:

$$W = R\int\left\{-\frac{a}{2}\mathbf{l}^2+\frac{1}{2}\mathbf{m}^2+\frac{c}{4}(\mathbf{lm})^2+\frac{k_1}{2}\left(|l_\perp|^2+|m_\perp|^2\right)-\frac{k_2}{4}\left(|l_\perp|^2+|m_\perp|^2\right)^2-\sqrt{k_1}hm_z+\right.$$
$$\left.+\frac{k_1}{2}\left[|\nabla l_\perp|^2+|\nabla m_\perp|^2+(\nabla l_z)^2+(\nabla m_z)^2\right]\right\}dxdydz \quad (2)$$

Here **m** and **l** are non-dimensional ferromagnetism and antiferromagnetism vectors; $l_\perp = l_x + il_y, m_\perp = m_x + im_y$; the absolute value of the vector **l** at $H=0$ equals $1$, $M_0$ is the magnetization of each sublattice, $x = K_1^{0.5}\alpha_{xy}^{-0.5}X$, $y = K_1^{0.5}\alpha_{xy}^{-0.5}Y$, $z = K_1^{0.5}\alpha_z^{-0.5}Z$; $h = HB^{-0.5}K_1^{-0.5}$, $k_1 = K_1/B$, $k_2 = K_2/B$, $h = H_z/\sqrt{BK_1}$, $a = A/B$, $c = C/B$, $R = 2M_0B\alpha_{xy}\alpha_z^{0.5}K^{-1.5}$.

The equations of motion with dissipative terms (in the Gilbert form) for the **l** and **m** vectors, taking into account the energy dissipation, are

$$\frac{M_0\hbar}{\mu_B}\frac{\partial\mathbf{l}}{\partial t} = \mathbf{m}\times\frac{\delta W}{\delta\mathbf{l}}+\mathbf{l}\times\frac{\delta W}{\delta\mathbf{m}}+\Gamma\frac{M_0\hbar}{\mu_B}\left(\mathbf{m}\times\frac{\partial\mathbf{l}}{\partial t}+\mathbf{l}\times\frac{\partial\mathbf{m}}{\partial t}\right) \quad (3)$$

$$\frac{M_0\hbar}{\mu_B}\frac{\partial\mathbf{m}}{\partial t} = \mathbf{m}\times\frac{\delta W}{\delta\mathbf{m}}+\mathbf{l}\times\frac{\delta W}{\delta\mathbf{l}}+\Gamma\frac{M_0\hbar}{\mu_B}\left(\mathbf{m}\times\frac{\partial\mathbf{m}}{\partial t}+\mathbf{l}\times\frac{\partial\mathbf{l}}{\partial t}\right). \quad (4)$$

The solutions of equations (2) and (3) can be presented in the form:

$$l_\perp(\mathbf{r},\tau) = q(\mathbf{r},\tau)e^{i(\omega(\tau)\tau-k(\tau)x)},\quad m_\perp(\mathbf{r},\tau) = p(\mathbf{r},\tau)e^{i(\omega(\tau)\tau-k(\tau)x)}. \quad (5)$$

For the simplicity, it is supposed that the excitations advance along the $x$-axis. The time dependencies of the $\omega$ and $k$ values are necessary for analyzing the time evolution of PBS.

From Eqs. (2)–(5), we have the following system of equations:

$$k_1^{-0.5}(\omega+\Delta+h)q = -\left(1-\frac{k_2}{k_1}(q^2+p^2)+k^2\right)(m_zq+l_zp)+\frac{a}{k_1}(m_zq-l_zp)+$$
$$+\frac{1}{k_1}(m_zq-l_zp)+m_z\Delta q-q\Delta m_z+l_z\Delta p-p\Delta l_z-QA_1, \quad (6)$$

$$k_1^{-0.5}(\omega+\Delta+h)p = -\left(1-\frac{k_2}{k_1}(q^2+p^2)+k^2\right)(l_zq+m_zp)+$$
$$+l_z\Delta q-q\Delta l_z+m_z\Delta p-p\Delta m_z-QA_2, \quad (7)$$

$$\frac{\partial p}{\partial \tau}=\sqrt{k_1}\left[2k\left(l_z\frac{\partial q}{\partial x}+m_z\frac{\partial p}{\partial x}\right)-Q(\omega+\Delta)(l_z q+m_z p)\right] \tag{8}$$

$$\frac{\partial q}{\partial \tau}=\sqrt{k_1}\left[2k\left(m_z\frac{\partial q}{\partial x}+l_z\frac{\partial p}{\partial x}\right)-Q(\omega+\Delta)(m_z q+l_z p)\right], \tag{9}$$

$$\frac{\partial l_z}{\partial \tau}=2\sqrt{k_1}\left[k\left(p\frac{\partial q}{\partial x}+q\frac{\partial p}{\partial x}\right)+Q(\omega+\Delta)qp\right], \tag{10}$$

$$\frac{\partial m_z}{\partial \tau}=-\sqrt{k_1}\left[k\frac{d}{dx}\left(q^2+p^2\right)-Q(\omega+\Delta)\left(q^2+p^2\right)\right], \tag{11}$$

where $\Delta=\left(\frac{d\omega}{d\tau}\tau-\frac{dk}{d\tau}x\right)$ and

$$A_1=\left(q\frac{\partial m_z}{\partial \tau}-m_z\frac{\partial q}{\partial \tau}+p\frac{\partial l_z}{\partial \tau}-l_z\frac{\partial p}{\partial \tau}\right), \tag{12}$$

$$A_2=\left(q\frac{\partial l_z}{\partial \tau}-l_z\frac{\partial q}{\partial \tau}+p\frac{\partial m_z}{\partial \tau}-m_z\frac{\partial p}{\partial \tau}\right). \tag{13}$$

In (6)–(13), the differentiation is carried out with respect to the dimensionless time $\tau=2\mu_B\left(K_1B\right)^{0.5}\hbar^{-1}t$; $Q=\Gamma 2\mu_B k_1^{-0.5}\hbar^{-1}$.

First of all, take notice that inserting Eqs. (8)–(11) into Eqs. (6) and (7), we can obtain a system of two modified equations describing the soliton configuration but not containing the evident time dependency. However, the time dependency of the parameters $p, q, l_z$ and $m_z$ characterizing the soliton form, described by Eqs. (8)–(11), is remaining. It means that during the soliton evolution connected with energy dissipation not only the configuration of PBS but also the frequency of precession $\omega$ and $k$ value are changing too. For each time moment, the definite configuration of PBS, frequency $\omega$ and wave value $k$ correspond. The correspondence between the changing values of $\omega$, $k$ and configuration is explicitly defined by the modified Eqs. (6) – (7).

To analyze the approximate soliton solutions of (6)–(11) system, it is convenient to transform the system of (6)-(11) equations into one equation. First of all, we will receive expression for magnetization. The relaxation time of ferromagnetic moment is less for some orders than the relaxation time for the antiferromagnetism vector. Therefore, for $m_Z$ component we use its quasi-equilibrium value, which can be obtained from $\frac{\delta W}{\delta m_Z}=0$ equation (see [8])

$$m_z\cong m_{z0}=\frac{1}{B+Cl_z^2}\left[H_z-\frac{C}{2}l_z\left(l_\perp m_\perp^*+l_\perp^* m_\perp\right)\right]=\frac{\delta\sqrt{k_1}h-(1-\delta)l_z pq}{l_z^2+\delta\left(1-l_z^2\right)}\cong-\frac{pq}{l_z}+\frac{\delta}{l_z^2}\left(\sqrt{k_1}h+\frac{pq}{l_z}\right) \tag{14}$$

where $\delta = B/(C+B) = \chi_{\parallel}/\chi_{\perp}$ is the ratio of two magnetic susceptibilities.

Taking into consideration that $k_1 << 1$, $k_2(q^2+p^2) << 1$ in the (6) equation and for long-wave oscillations $k_1\Delta q << q$, $k_1\Delta l_z << l_z$, $k_1\Delta m_z << m_z$, $k_1\Delta p << p$, from the (6) and (14) equations we obtain the following dependence

$$p \cong -\frac{\sqrt{k_1}(\omega+\Delta+h)ql_z}{q^2+l_z^2}\left[1-a\left(2q^2-1\right)\right] + \delta(1+a)\sqrt{k_1}\frac{q}{l_z}\left(h-(\omega+\Delta+h)q^2\right) - \\ - k_1\sqrt{k_1}k^2(\omega+\Delta+h)ql_z\left(2q^2-1\right) - Ql_zA_1 \tag{15}$$

Using the (15) correlation, from (7) we obtain the following equation for PBS:

$$\Delta q - \frac{q}{l_z}\Delta l_z = \left[1-(1+a)(\omega+\Delta+h)^2 + k^2 + \delta\frac{(1+a)(\omega+\Delta+h)h}{l_z^2}\right]q - \\ -\left[\frac{k_2}{k_1} - 2a(\omega+\Delta+h)^2 + 2k_1(\omega+\Delta+h)^2k^2 + \delta(1+a)\frac{(\omega+\Delta+h)^2}{l_z^2}\right]q^3 + Q\frac{A_2}{l_z} \tag{16}$$

Let's confine ourselves to low temperature approximation, i.e. suppose that $(\mathbf{lm}) = 0$. In such case $a = \delta = 0$, $q^2+p^2+l_z^2+m_z^2 = 1$,

$$A_1 = 2\sqrt{k_1}k\left[3qp\frac{\partial q}{\partial x} + \left(2q^2+p^2-1\right)\frac{\partial p}{\partial x} - \frac{1}{2}p\frac{d}{dx}\left(q^2+p^2\right)\right] + Q(\omega+\Delta)p, \tag{17}$$

$$A_2 = 2\sqrt{k_1}k\left[3pq\frac{\partial p}{\partial x} + \left(2p^2+q^2-1\right)\frac{\partial q}{\partial x} - \frac{1}{2}q\frac{d}{dx}\left(q^2+p^2\right)\right] + Q(\omega+\Delta)q \tag{18}$$

and the equation for PBS is as follows:

$$\Delta q - \frac{q}{l_z}\Delta l_z = \left(1-(\omega+\Delta+h)^2+k^2\right)q - \left(\frac{k_2}{k_1} + 2k_1(\omega+\Delta+h)^2k^2\right)q^3 + Q\frac{A_2}{l_z}. \tag{19}$$

This equation should be supplemented by the following expression obtained from (9), (14), (15) at $a = \delta = 0$ and describing the evolution of PBS:

$$\frac{\partial q}{\partial \tau} \cong k_1(\omega+\Delta+h)\left[2k\left(3q^2-1\right)\frac{\partial q}{\partial x} - Q(\omega+\Delta)q\left(2q^2-1\right)\right]. \tag{20}$$

Only in the case of immovable PBS, i.e. at $k \equiv 0$, the (19) equation has the solutions with a spherical symmetry. For such case, the equation has the following form:

$$\frac{d^2q}{dr^2} + \frac{2}{r}\frac{dq}{dr} + \frac{q}{1-q^2}\left(\frac{dq}{dr}\right)^2 = q\left[\left(1-q^2\right)\left(1-(\omega+h)^2 - \frac{k_2}{k_1}q^2\right) + Q^2\omega\left(1+\frac{1}{2}q^2\right)\right] \tag{21}$$

(taking into account that $l_z^2 \cong 1-q^2$). In this equation, the frequency $\omega$, as well as $q$ values, depends on time. Thus we make the replacement $\left(\omega + \frac{d\omega}{d\tau}\tau\right) \to \omega(\tau)$, which is possible considering a rather slow change of the precession frequency in comparison with the change of precession phase. Note that in the (21) equation, the addition connected with the dissipation is negligibly small, $Q^2|\omega| << 1$ (in our examples $Q = 0.02, |\omega| < 0.03$). Therefore, in further calculations we will neglect this addition.

Transforming Eq. (21) with respect to $l_z$ parameter, we can obtain the following equation, that can be used to analyze the PBS originated during the reverse phase transition, at decrease of the field:

$$\frac{d^2 l_z}{dr^2} + \frac{2}{r}\frac{dl_z}{dr} + \frac{l_z}{1-l_z^2}\left(\frac{dl_z}{dr}\right)^2 = \left((\omega+h)^2 - 1 + \frac{k_2}{k_1}\right)l_z - \left((\omega+h)^2 - 1 + \frac{2k_2}{k_1}\right)l_z^3 + \frac{k_2}{k_1}l_z^5 . \quad (22)$$

Using Eqs. (14) and (15) at $\delta = 0, a = 0$ and $k_1 << 1$, from (2) we obtain the following expression for the energy of PBS:

$$E_s = 8\pi M_0 \alpha_{xy}\sqrt{\frac{\alpha_z}{k_1 B}}\int_0^\infty \left\{\left[\frac{1+(\omega+h)^2}{2} - (\omega+h)h\right]q^2 - \frac{k_2}{4k_1}q^4 + \frac{1}{2(1-q^2)}\left(\frac{dq}{dr}\right)^2 + \frac{k^2}{2}q^2\right\}r^2 dr . \quad (23)$$

The last term in this integral corresponds to the kinetic energy of PBS.

## 2. Spontaneous origin of PBS and their evolution into domains of a new phase

As in [8], we divide the processes relating to PBS into two stages. At the first stage, the PBS originate spontaneously because of the thermal fluctuations, or by any different way, in the $\tau = 0$ moment, and further evolution of PBS is carried out at the second stage. Equations (21) or (22) are used for description of the form of arose PBS. At this stage, we do not take into account the dependence of PBS parameters on the time and dissipation of energy.

PBS configurations, that are particular solutions of Eq. (21), for different $\omega$ values at $h = 0.99$, and dependences of PBS energy and amplitudes on the precession frequency for different values of the field have been shown in [8]. In Fig.1, we present the frequency dependencies for the energy, amplitude ($q_m \equiv q(r=0)$) and effective radius for PBS at $h = 0.99$ only. In this case, the utmost frequency of PBS $\omega = 0.01$ corresponds to the frequency of antiferromagnetic resonance, $\omega_{res1} = (1-h)$. In the Fig.2, we can see the same dependencies for the reverse spin-flop transition (at the decreasing magnetic field) at $h = 0.91$. In this case, the frequency of antiferromagnetic resonance can be expressed as follows: $\omega_{res2} = -h + \sqrt{1 - k_2/k_1} \cong -0.01557$. Here and in all subsequent examples the following parameters typical for $Cr_2O_3$ have been used: $M_0 = 0.33 \times 10^{-9} eVOe^{-1}Å^{-3}$, $B = 4.9 \times 10^6 Oe$, $\alpha_{xy} = \alpha_z = 3 \times 10^6 Oe\,Å^2$, $K_1 = 700\, Oe$, $K_2 = 140\, Oe$, $Q = 0.02$ [9].

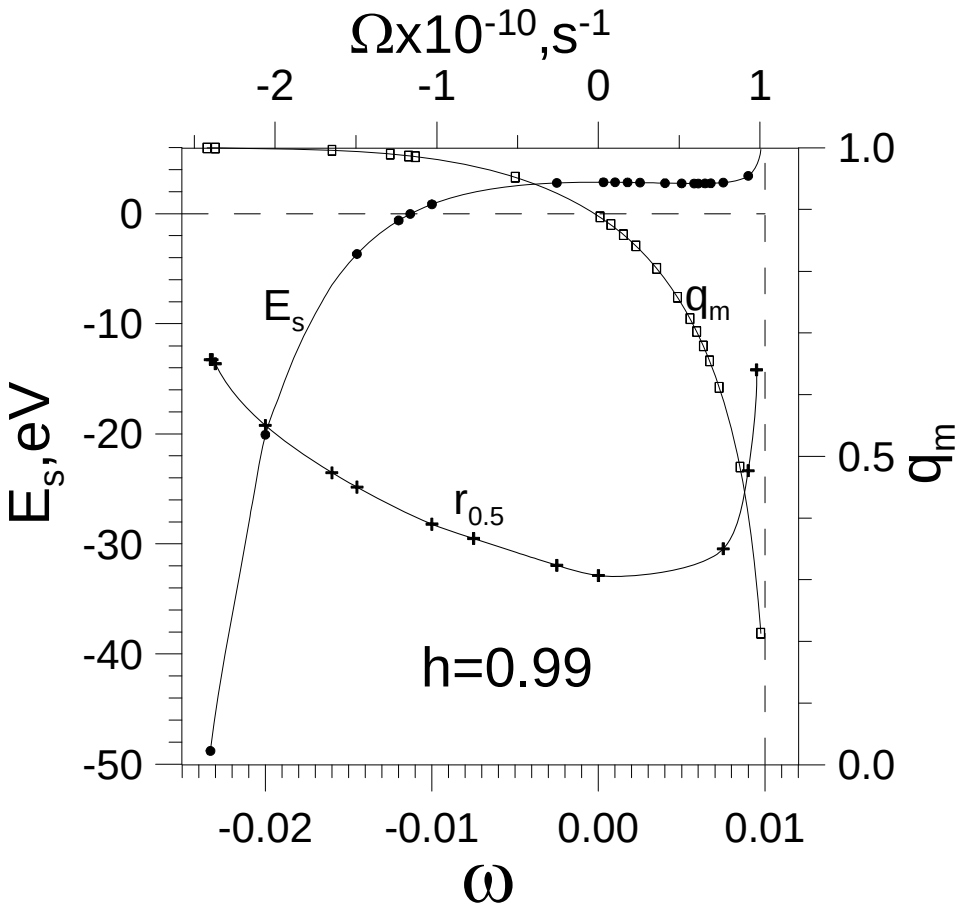

Fig.1 Frequency dependencies of the energy, amplitude and effective radius if $h = 0.99$. Here and in all subsequent Figures, the values of energy are denoted by circles, the amplitude – by empty squares, the frequency and the inverse of the frequency – by continuous line or full squares, the radius – by crosses.

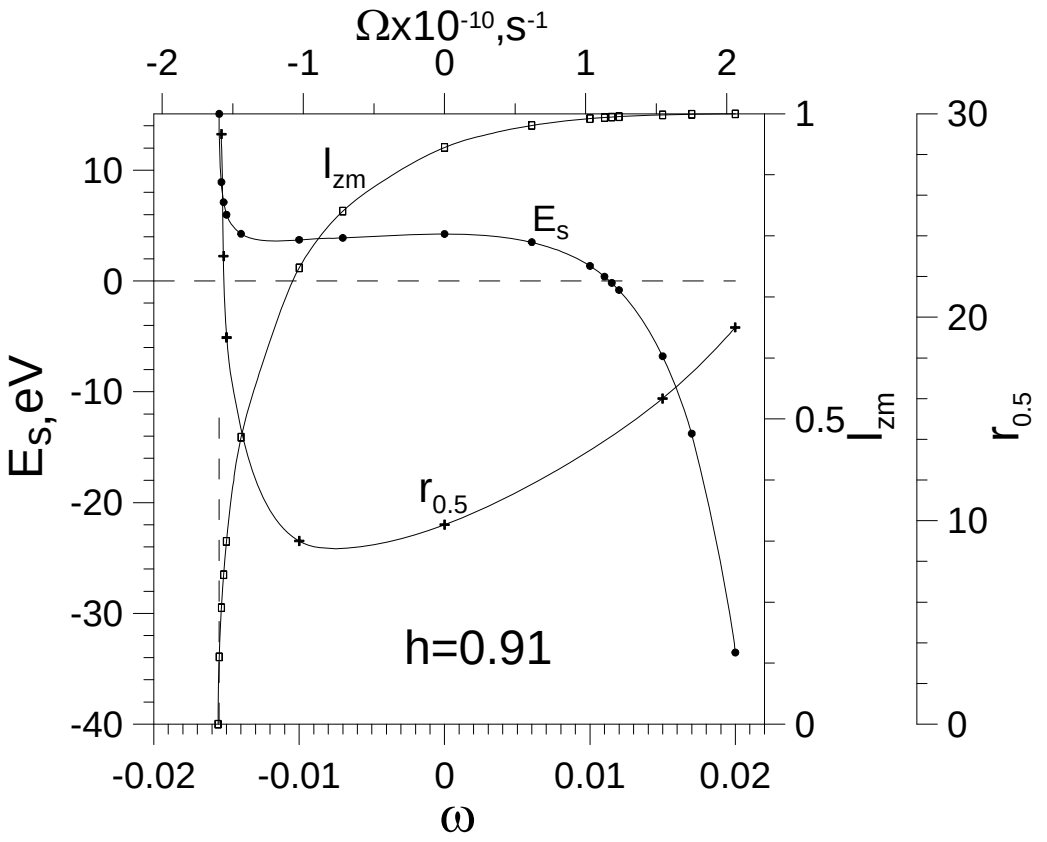

Fig.2 The same frequency dependencies as in Fig.1 for the reverse spin-flop transition, i.e. at the decreasing magnetic field, if $h = 0.91$.

Probability of the spontaneous creation of PBS related to the fluctuations of energy at non-zero temperatures is proportional to probability of such fluctuations. We can use the following expression for the probability of PBS creation with the energy $E_S$ near the bifurcation point (see [7, 8]):

$$P_s = A_s \exp\left(\frac{-|E_s|}{k_B T}\right). \tag{24}$$

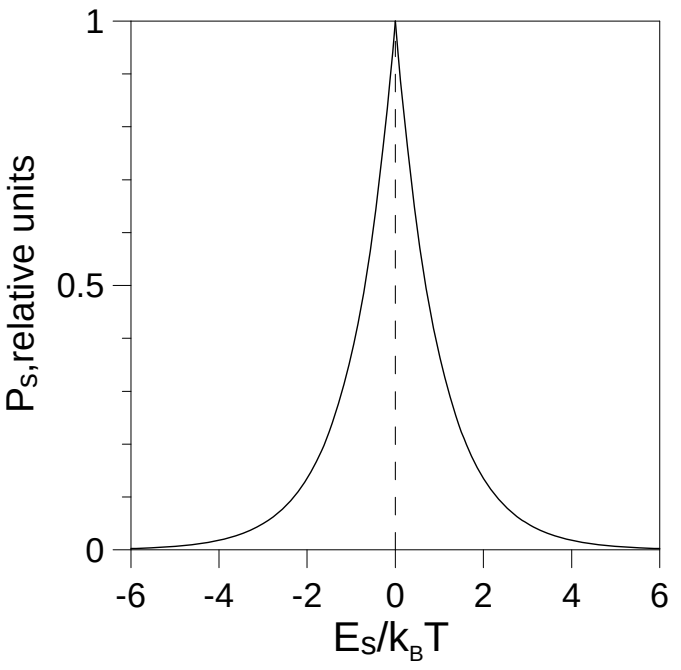

Fig.3 Energy dependence for probability of the spontaneous PBS creation near the point of bifurcation.

Here $A_s$ is the coefficient in general depending on ω, $E_s$ and configuration of PBS. Generally, we have to use different configuration coefficients for positive and negative energies of PBS: $A_{s+}$ for $E_s > 0$ and $A_{s-}$ for $E_s < 0$. Apparently that $A_{s+}, A_{s-} << 1$. In Fig.3, the temperature dependence of the probability of PBS near to the bifurcation point, where the energy of PBS is near to zero, is shown.

The further evolution of PBS is described by Eq. (20). As it can be seen from this equation, subsequent change of PBS configuration depends on two factors: the spatial movement of it as a whole and dissipation of energy.

In this part, we consider the change of PBS related to the dissipation only, i.e. at $k \equiv 0$. Let's assume that at the moment $\tau = 0$ the PBS of the $q(r,0)$ form, corresponding to Eq. (21), have arisen. A subsequent change of PBS configuration proceeds in accordance with expression

$$\frac{\partial q}{\partial \tau} \cong -Qk_1\omega(\omega + h)q(2q^2 - 1) \tag{25}$$

where $\omega = \omega(\tau)$.

In conformity with Eq.(25), the character of PBS change is determined by a sign of the precession frequency and by its amplitude.

In Fig.4 – Fig.7, several examples are adduced at $h < 1$ to illustrate the PBS change because of the dissipation for different field values. If the amplitude of PBS is big enough, we can consider it as consisting of two parts: the "bulk", where value $q > \sqrt{0.5}$, and the "corona", where $q < \sqrt{0.5}$.

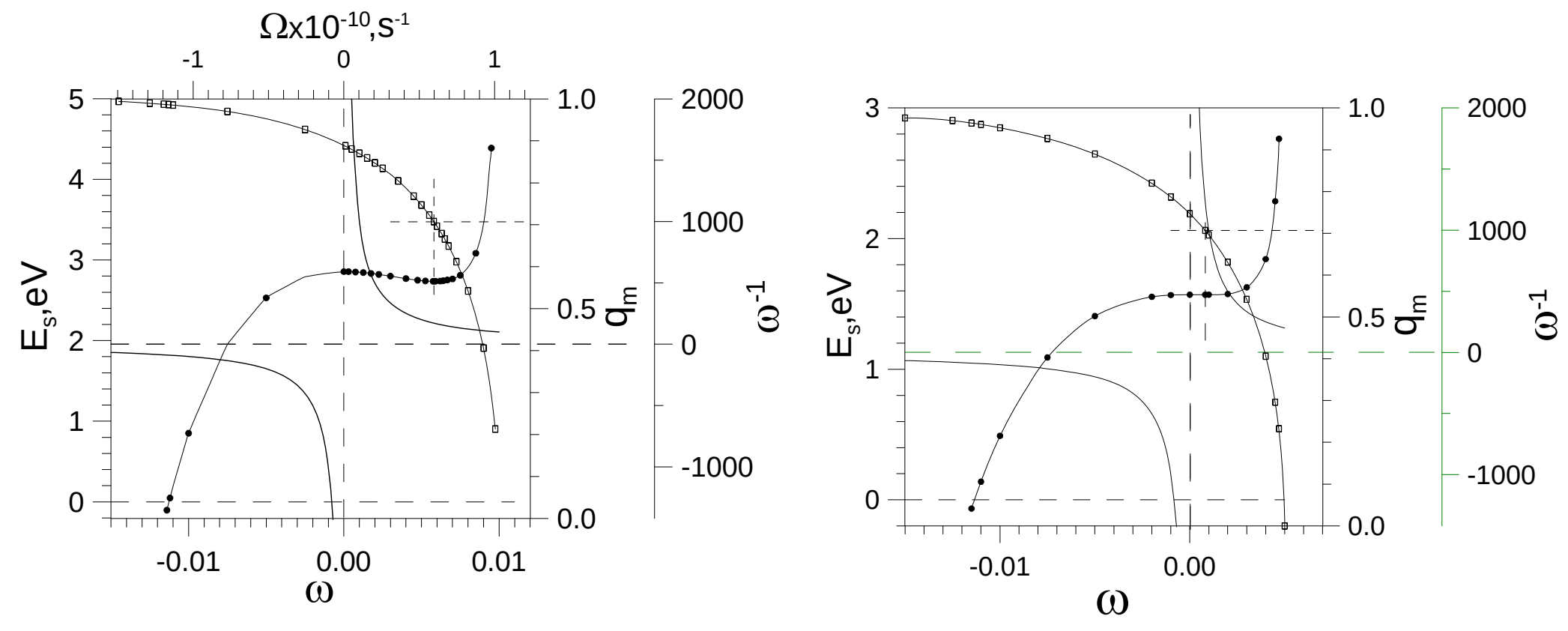

Fig.4 The frequency dependencies of energy, amplitude and $\omega^{-1}$ value if $h = 0.99$.

Fig.5 The same frequency dependencies as in Fig.4 if $h = 0.995$.

If $\omega_{init} < 0$ and the amplitude of PBS $q_{m,init} > \sqrt{0.5}$, its bulk is increasing, but the corona is decreasing. As a result, the value of $r$ corresponding to $q = \sqrt{0.5}$ is increasing, the frequency ω is increasing in absolute value, the energy is decreasing, PBS is growing and turning into the macroscopic domain of a high-field phase. Changes of PBS can be seen at $\omega_{init} < 0$ in Fig.4 and in Fig.5, at $\omega_{init} < -0.0011$ in Fig.6, and at $\omega_{init} < -0.00419$ in Fig.7. On the contrary, if $\omega_{init} > 0$ and $q_{m,init} < \sqrt{0.5}$, the PBS is decreasing in amplitude. Such change can be traced in Fig.4 in the range $0 < \omega < 0.0058$ and in Fig.5 in the range $0 < \omega < 0.0008$.

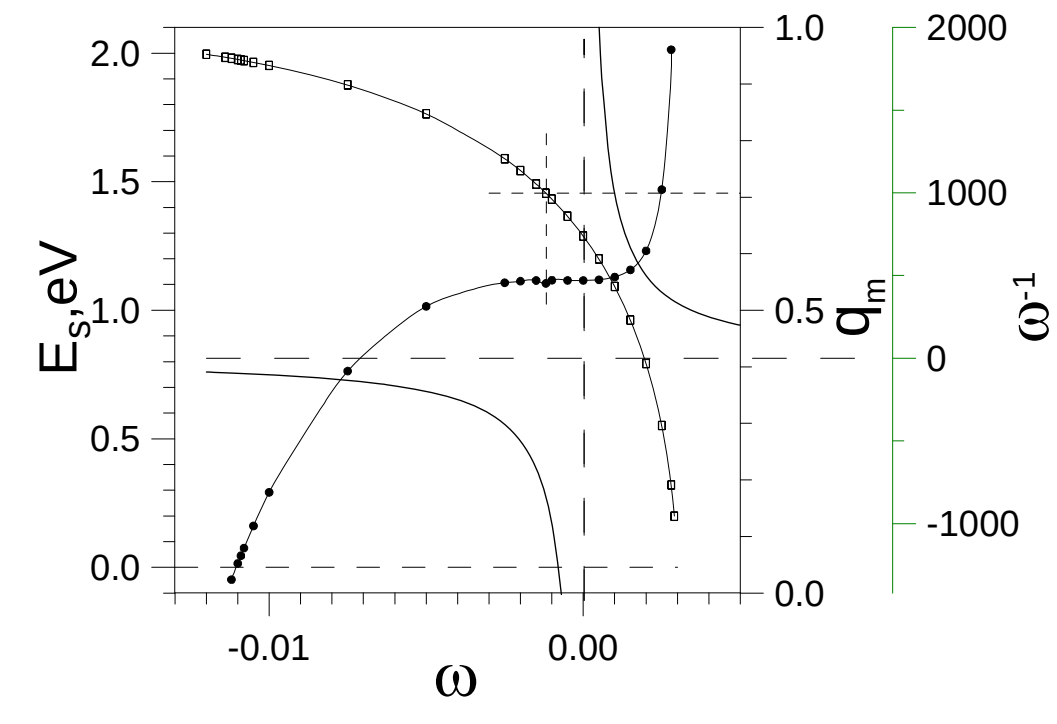

Fig.6 The same dependencies as in Fig.4 if $h = 0.997$.

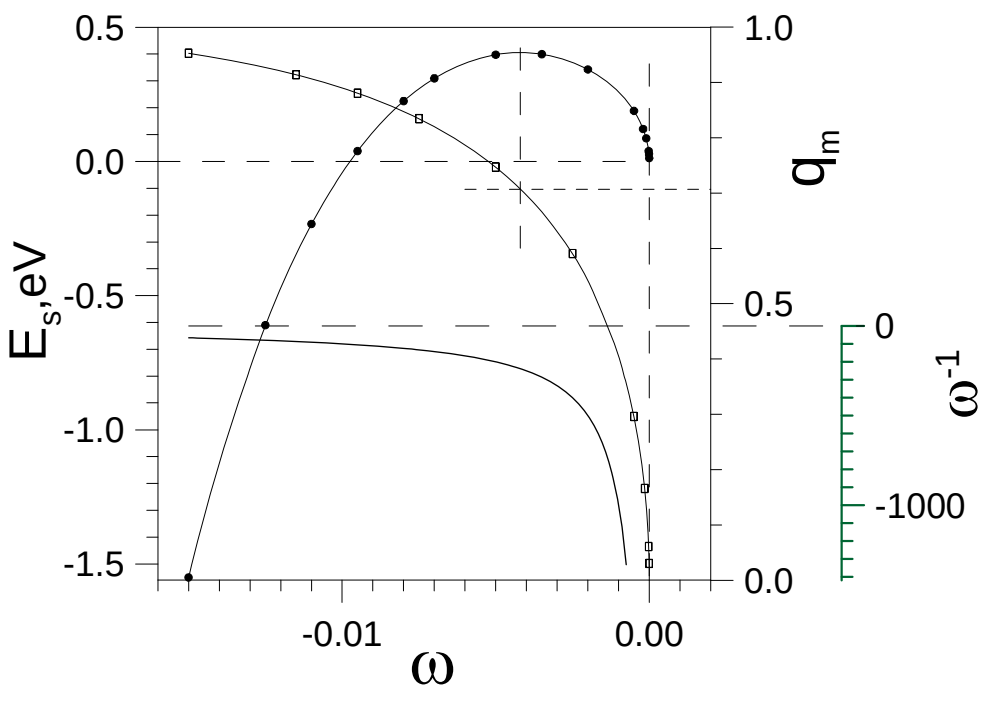

Fig.7 The same dependencies as in Fig.4 if $h = 1$

If $\omega_{init} < 0, q_{m,init} < \sqrt{0.5}$, the $q$ values in the bulk and corona of PBS are decreasing (see, for example, at $0 > \omega > -0.00119$ in Fig.6, and at $\omega > -0.00419$ in Fig.7). At the same time, the frequency ω is decreasing in absolute value, and $\omega \to -0$. In the last case, at $h = 1$, the PBS disappear asymptotically, i.e. $q_m \to 0, E_s \to 0$.

Thus, only in cases $\omega_{init} < 0,\ q_{m,init} > \sqrt{0.5}$ PBS are transforming into macroscopic domains of the high-field phase.

Equations (21) and (25) permit to obtain the time dependence of the frequency (and energy) due the dissipation, in accordance with the succession: $\Delta\omega \to \Delta q_m \to \Delta\tau \to \omega(\tau)$. For each change of the frequency $\Delta\omega$, from Eq. (21) we obtain the corresponding change of PBS amplitude $\Delta q_m$; further from Eq.(25), we obtain the time interval $\Delta\tau$ corresponding to this change of amplitude. As a result, we obtain the $\omega(t)$ dependences for the cases of transformation of PBS into domains of a new phase. They are presented in Fig.8 – Fig.11.

Figures 8 and 9 correspond to the field $h = 0.99$. Initial data for PBS in Fig.8 are $\omega_{init} = -0.0025,\ E_{init} \cong 2.8553 eV$, and in Fig.9: $\omega_{init} = -0.0112,\ E_{init} \cong 0.0476 eV$. Note that at $T = 300K$ the average energy of the thermal fluctuations equals $k_B T = 0.025 eV$. The frequencies corresponding to value $E_s \cong 0$ are marked by dotted lines. Practically, the curve in Fig.9 is a part of the Fig.8 curve.

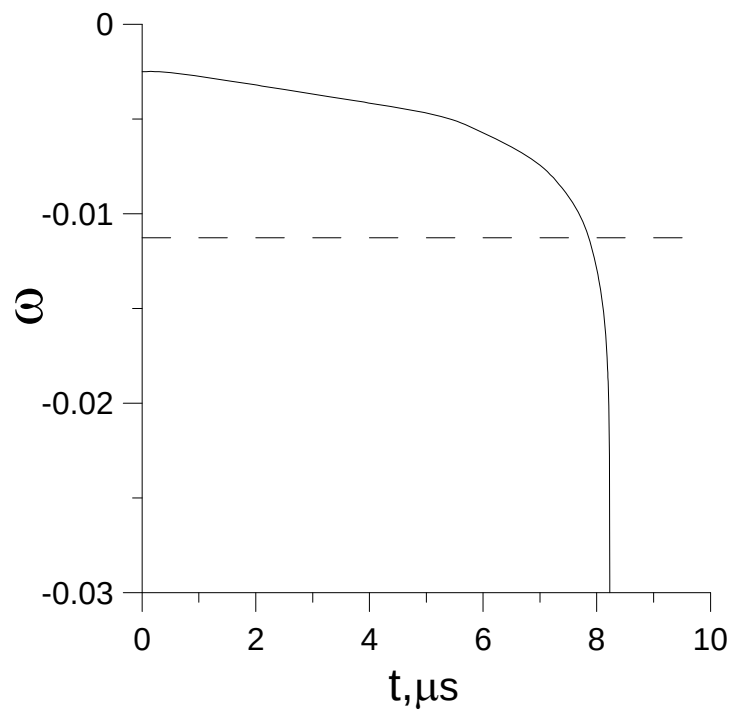

Fig.8 The time dependency of the frequency if $h = 0.99$. Here the initial value of energy is $E_{init} \cong 2.855\, eV$ (see in Fig.4). The dotted line shows the frequency value corresponding to zero energy.

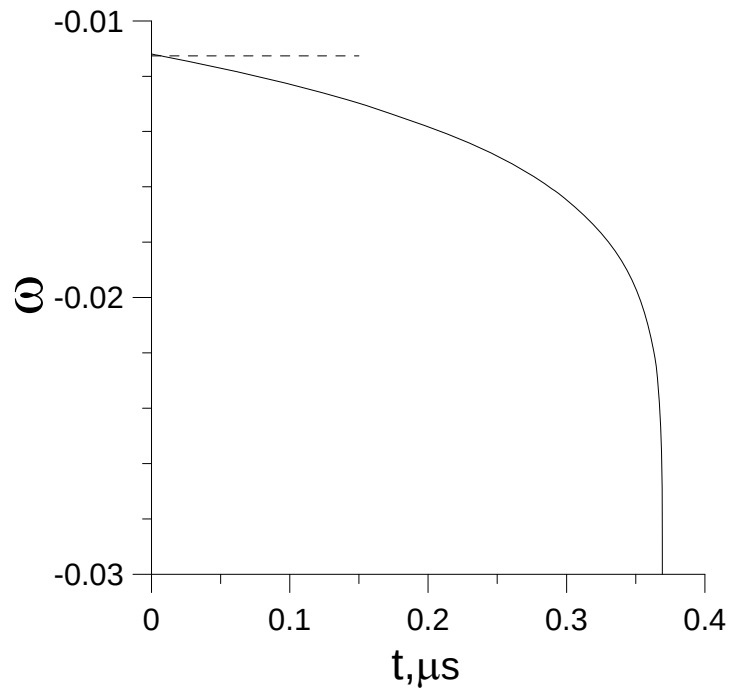

Fig.9 Time dependency of the frequency if $h = 0.99$, but here $E_{init} \cong 0.0476\, eV$.

Figure 10 corresponds to the field $h = 0.997$, and in this case $E_{init} = 0.2919 eV$. At last, Fig.11 corresponds to $h = 1$, $E_{init} = 0.3977 eV$.

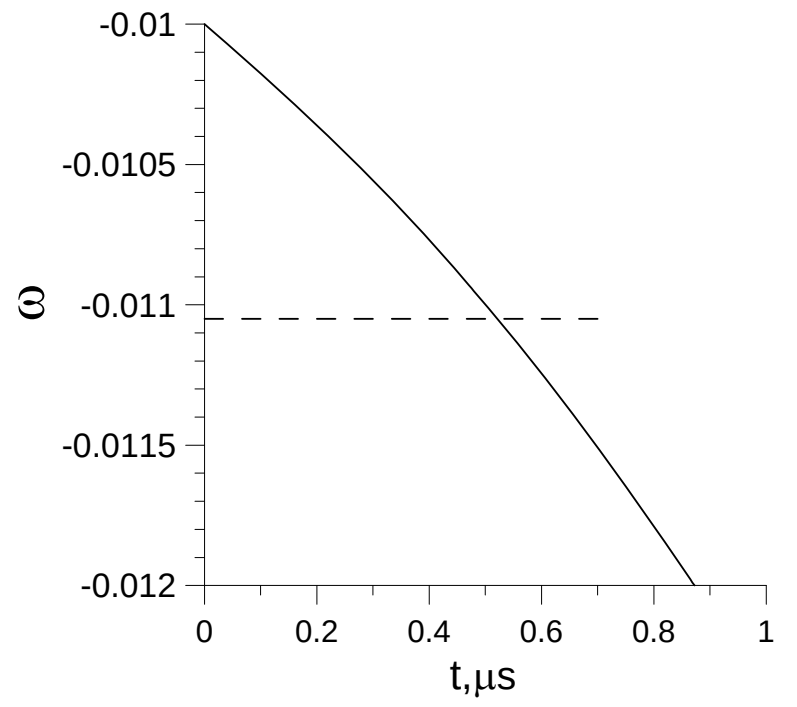

Fig.10 The time dependency of the frequency if $h = 0.997$. Here $E_{init} \cong 0.292 eV$ (see in Fig.6).

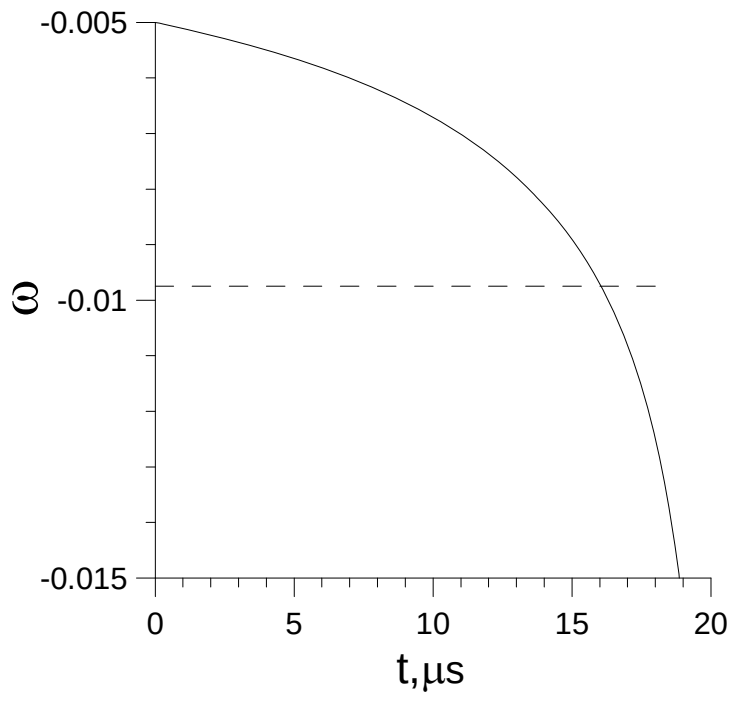

Fig.11 Time dependency of the frequency if $h = 1$ and the initial amplitude $q_{m,init} > \sqrt{0.5}$. Here the initial parameters of PBS are the following: $E_{init} \cong 0.398\, eV$, $q_{m,init} = 0.747$, $\omega_{init} = -0.005$ (see in Fig.7).

Sequence of change of the form and sizes for PBS at $h = 0.99$, $\omega_{init} = -0.0025$ has been shown in Fig.12. (Note that the similar sequences presented in [8] are wrong, since the frequency change connected with dissipation has not been taken into consideration.)

Asymptotical disappearance of PBS at $h = 1$ is illustrated in Fig.13.

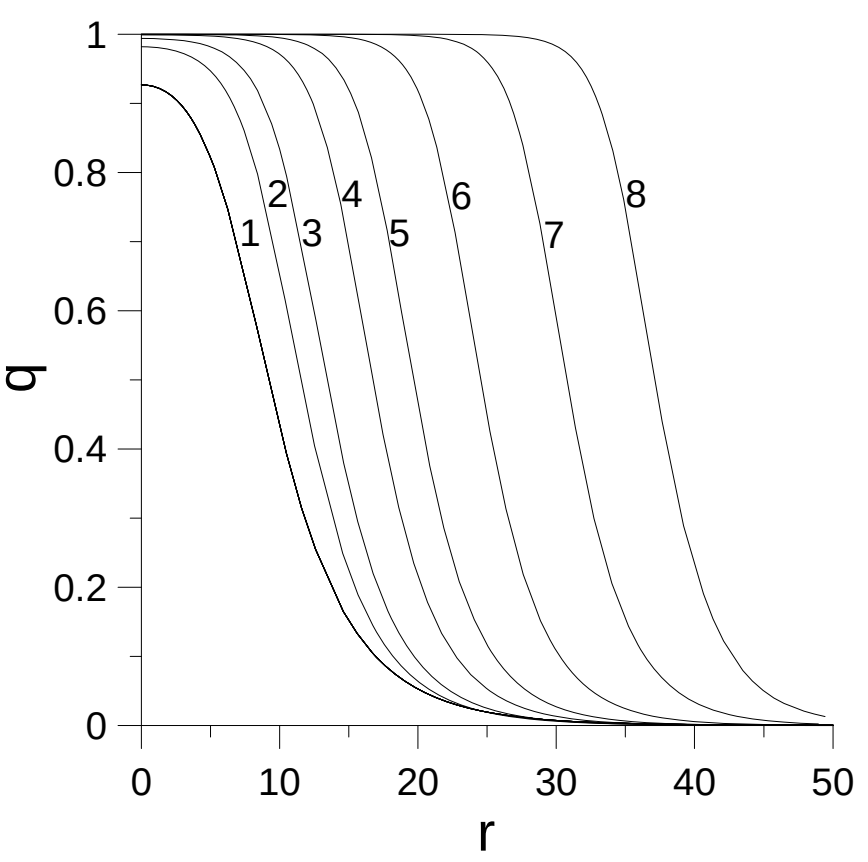

Fig.12 A sequence of configurations of PBS if $h = 0.99$ and the initial amplitude $q_{m,init} > \sqrt{0.5}$ corresponding to the Table.

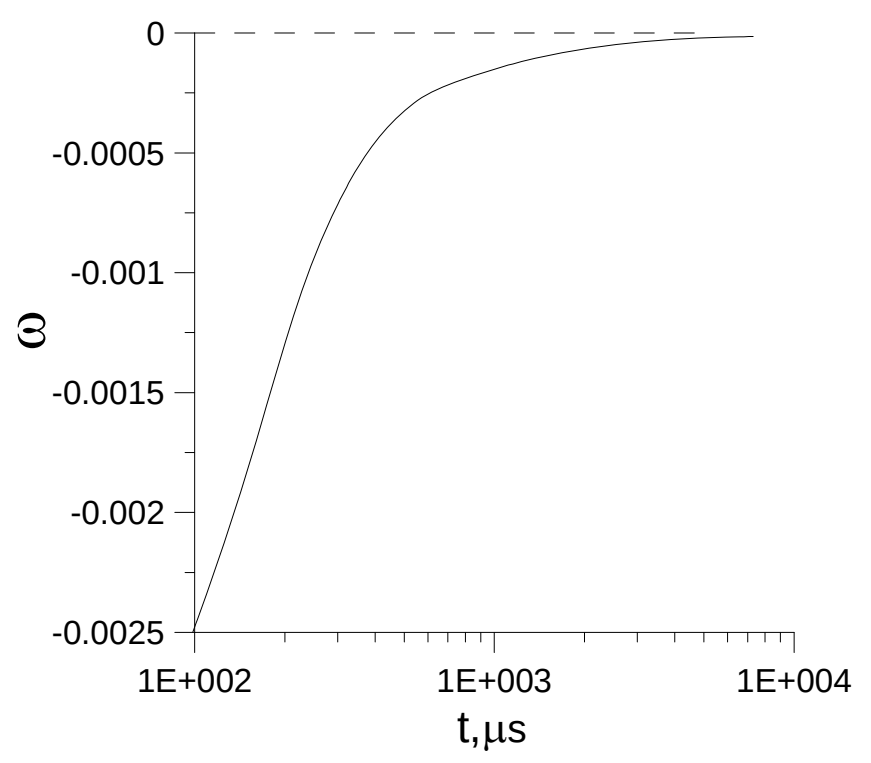

Fig.13 Damping (fading) of PBS if $h = 1$ and the initial amplitude $q_{m,init} < \sqrt{0.5}$. Here the parameters of initial PBS are the following: $\omega_{init} = -0.0035$, $q_{m,init} \cong 0.666$, $E_{init} = 0.3996\, eV$ (maximum energy of PBS for $h = 1$ equals $E_{s\max} = 0.405465\, eV$, see in Fig.7).

Table

| № | t (μs) | ω | $E_s$ (eV) | $R_{0.5}$ (Å) |
|---|---|---|---|---|
| 1 | 0 | -0.0025 | 2.7892 | 635 |
| 2 | 7.6896 | -0.01 | 0.8513 | 765 |
| 3 | 8.0899 | -0.0145 | -3.6714 | 885 |
| 4 | 8.2097 | -0.02 | -20.08 | 1080 |
| 5 | 8.2236 | -0.0233 | -48.8 | 1290 |
| 6 | 8.2243 | -0.027 | -130.7 | 1610 |
| 7 | 8.2263 | -0.03 | -321.23 | 2000 |
| 8 | | -0.032 | -791.8 | 2580 |

## 3. Equilibrium ball solitons

At $h \le 1$, the curves of $E_s(\omega)$ have a minimum as it is obviously in Fig.14. Let's consider the regions of the minima only in two cases: at $h = 0.99$ in Fig.15 and $h = 0.999$ in Fig.16. In the first case, the energy minimum is at $\omega \cong 0.0058, q_m = \sqrt{0.5}$, and the energy maximum is at $\omega = 0$. In the second case, the minimum is at $\omega = 0$ and the maximum at $q_m = \sqrt{0.5}$.

Asymptotical approach of precession frequency to equilibrium values as the time function to $\omega \cong 0.0058$ in the first case and to $\omega = 0$ in the second case, on the side of lesser and greater values, are shown in Fig.17 and Fig.18.

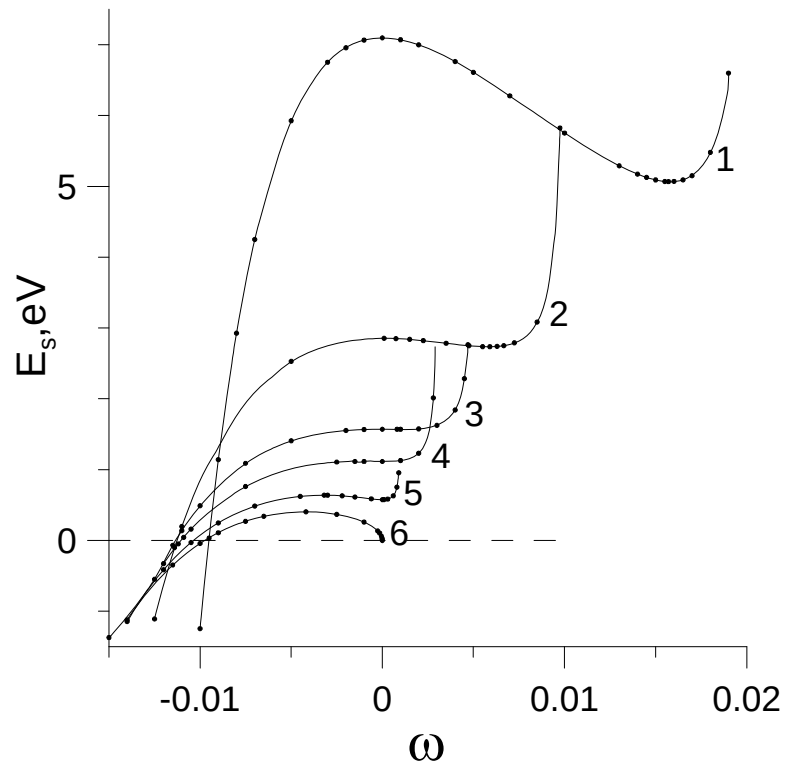

Fig.14 Frequency dependencies of PBS energy for the following values of a field: (1) – 0.98; (2) – 0.99; (3) – 0.995; (4) – 0.997; (5) – 0.999; (6) - 1.

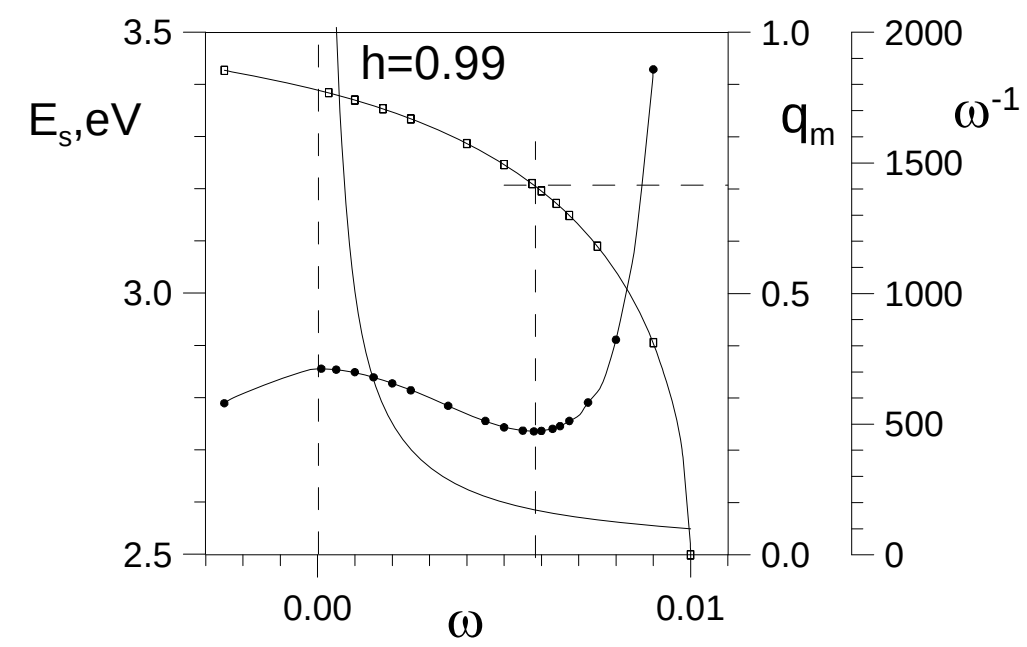

Fig.15 Frequency dependencies of energy, amplitude, and $\omega^{-1}$ value for $h = 0.99$.

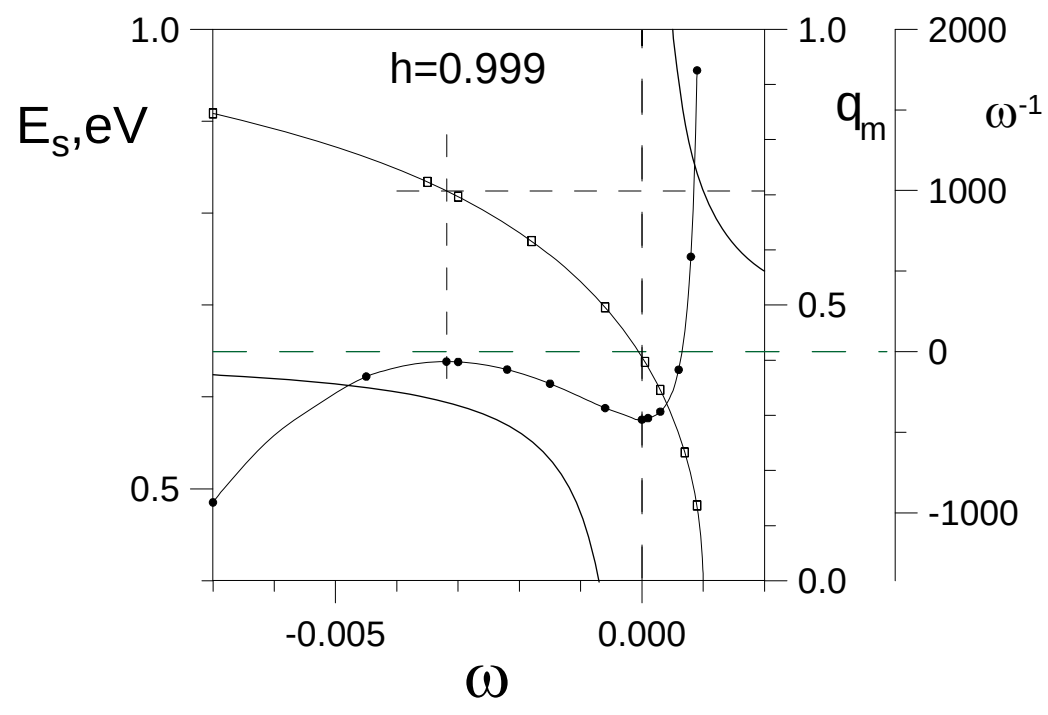

Fig.16 Frequency dependencies of energy, amplitude, and $\omega^{-1}$ value for $h = 0.999$.

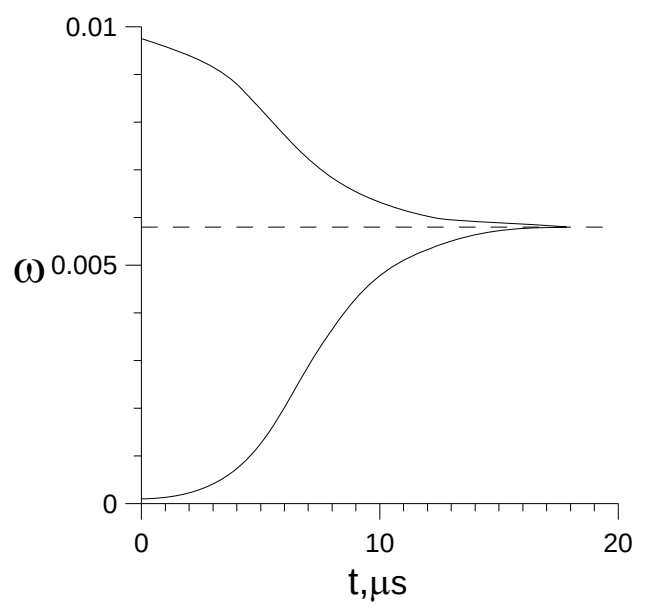

Fig.17 The change of the frequency at approaching of PBS to the equilibrium from the side of lesser and greater values of the frequency if $h = 0.99$. Equilibrium frequency is $\omega_{equal} \cong 0.0058$.

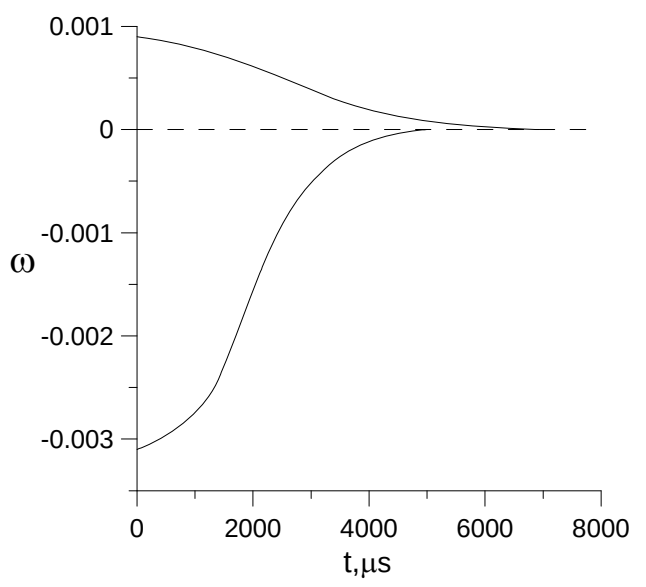

Fig.18 The change of the frequency at approaching of PBS to the equilibrium from the side of lesser and greater values of the frequency if $h = 0.999$.

Thus, at $h<1$ we have two types of equilibrium solitons: precessing equilibrium ball soliton of the first type (EBS-1), and non-precessing equilibrium solitons of the second type (EBS-2). EBS-1 have the fixed amplitude equal to $q_m = \sqrt{0.5}$ and the configuration that does not depend on the field values. The amplitude and configuration of EBS-2, at $\omega = 0$, depend on the magnetic field. The states of EBS-1 exist at $h < h^*$, in our examples $h^* \cong 0.996$. The states of EBS-2 correspond to condition $1 > h > h^*$.

In Fig.19, the field dependencies of the energy, amplitude and frequency for the equilibrium PBS are shown. In Fig.20, the configurations of equilibrium solitons are presented.

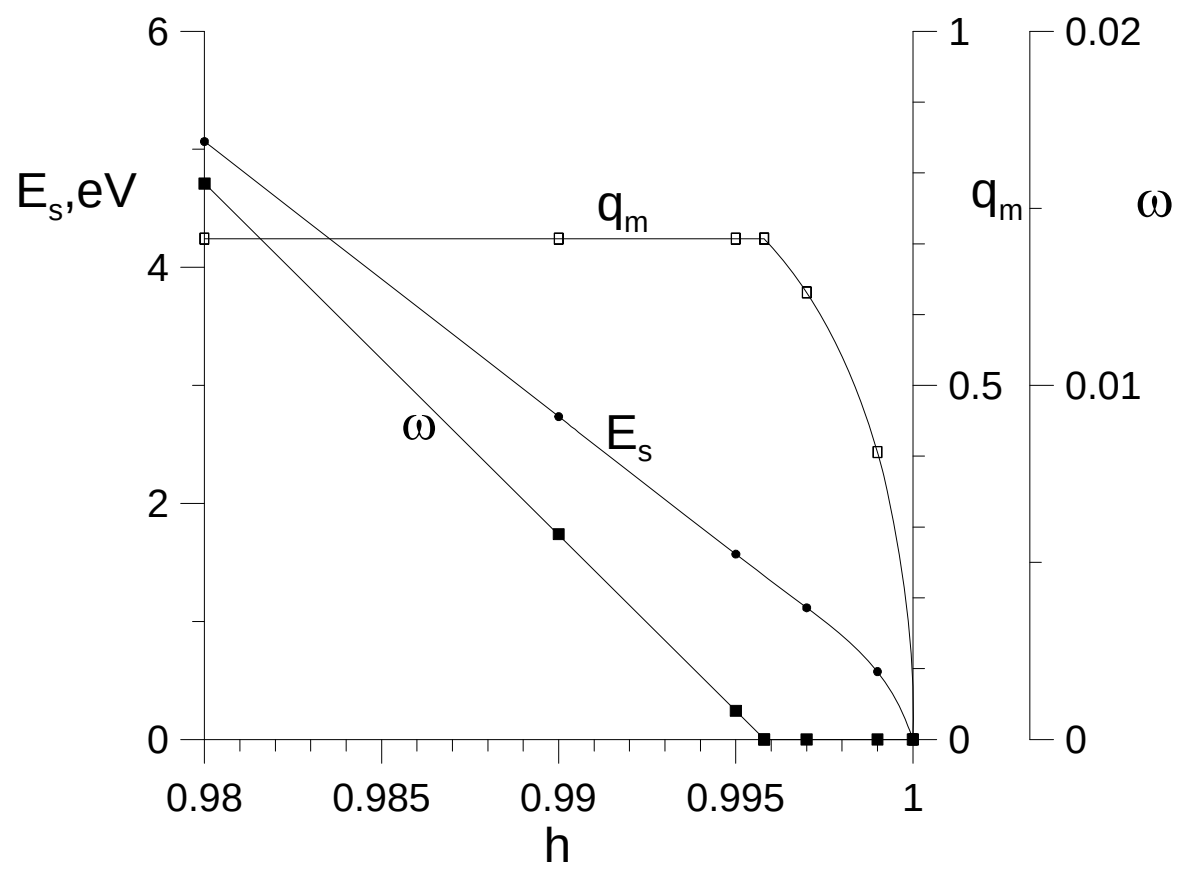

Fig.19 The field dependencies of the energy, amplitude, and frequency for the equilibrium PBS.

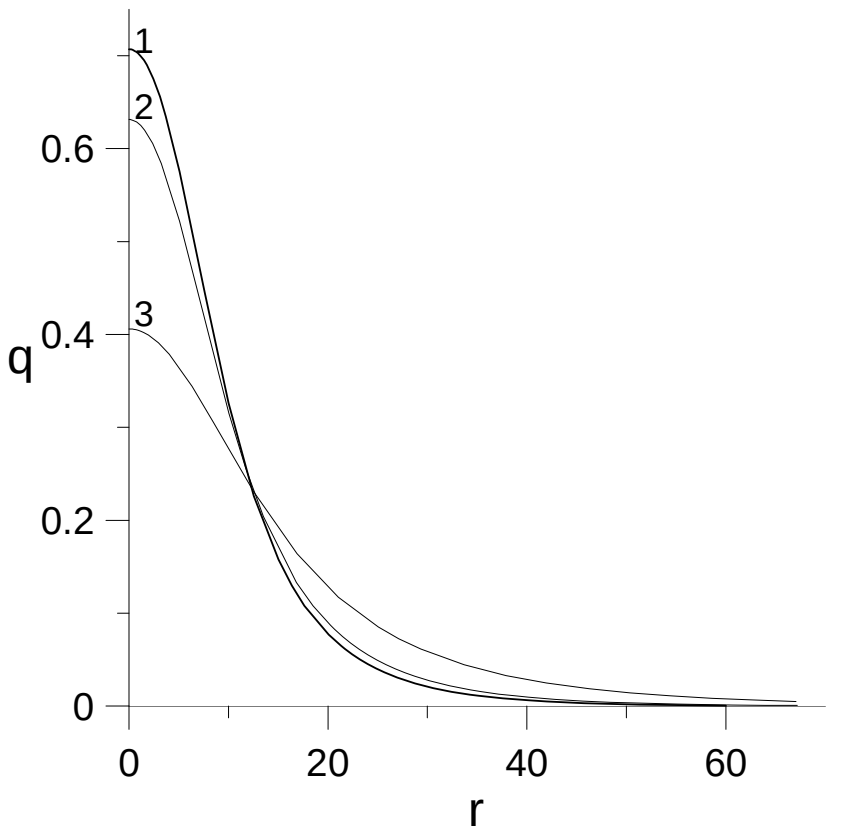

Fig.20 Configurations of the equilibrium PBS: (1) $q_m = \sqrt{0.5}$, (2) $q_m = 0.631$, $h = 0.997$, (3) $q_m = 0.406$, $h = 0.999$.

## 4. Overcritical PBS

The PBS states also exist when the initial phase state is absolutely unstable [6-8]. Therefore, creation of PBS is possible at disintegration of the initial phase, i.e. at $h > 1$ (and at $h < \sqrt{1 - k_2/k_1}$ for reverse spin-flop transition). Besides, there is the range of values of a magnetic field, where the energy of such PBS can be positive. It is visible in Fig.21, for example, that at $h = 1.001$ the creation of PBS, having the energy near to zero, is possible with two various frequencies of precession. It can be seen from Eq.(21) that $(\omega + h)^2 \le 1$, i.e. for all PBS at $h > 1$ the precession frequencies are negative.

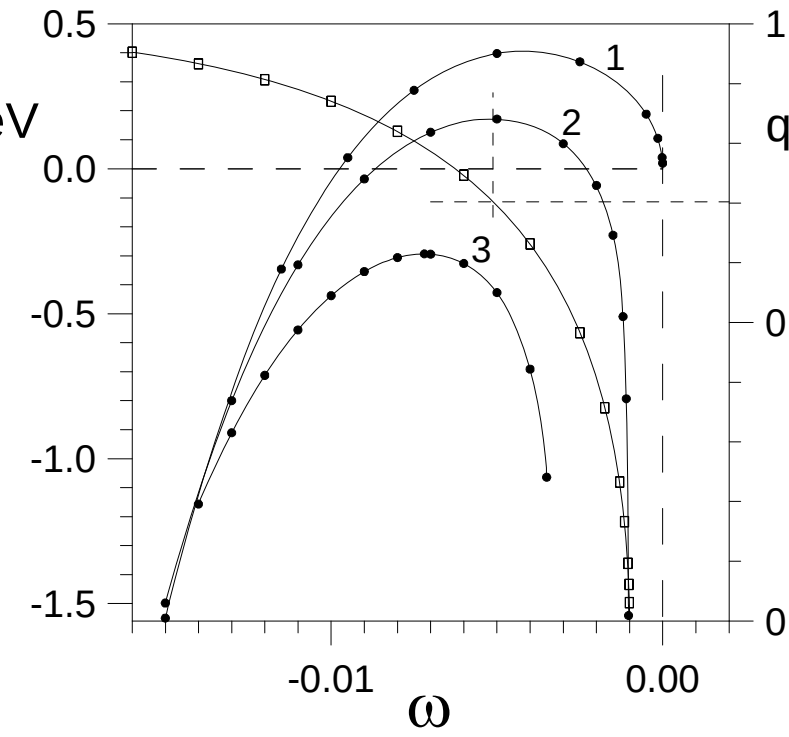

Fig.21 Frequency dependency of PBS energy for the overcritical values of a field: 1 - $h = 1$, 2 - $h = 1.001$, 3 - $h = 1.003$. The curve of amplitude is shown for $h = 1.001$.

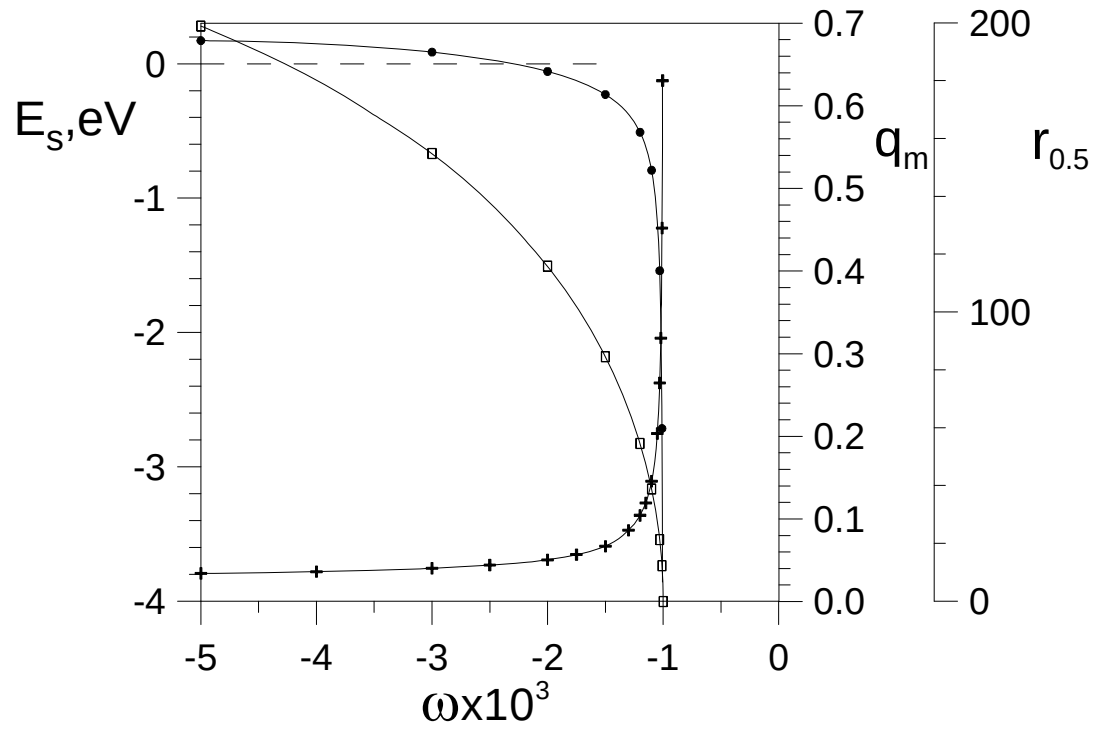

Fig.22 Frequency dependencies of energy, amplitude, and radius of PBS for $h = 1.001$, $q_{m,init} < \sqrt{0.5}$. Here $q_{m,init} = 0.696$.

In Fig.22, the frequency dependencies of energy, amplitude, and radius of PBS at $h = 1.001$, $q_m < \sqrt{0.5}$ (it corresponds to curve 2 in Fig.21) are shown. In this case, in correspondence with Eq.(25), the frequency $\omega$ is decreasing in absolute value and all $q$ values in PBS, for each radius, are decreasing, i.e. $\omega \to -0.001$ and $q_m \to 0$. However, the energy is getting negative and decreasing too. Such change of energy can be explained by the fact that simultaneously the radius of PBS is increasing considerably. The time dependencies of the main parameters of PBS in this case are presented in Fig.23. Thus, at $h$ value, exceeding the critical value a little, PBS of long duration can originate which are decreasing in amplitude and simultaneously increasing in radius. Of course, PBS of long duration exist only if they are considered in isolation from other processes. In reality, they will be absorbed by other more quick objects.

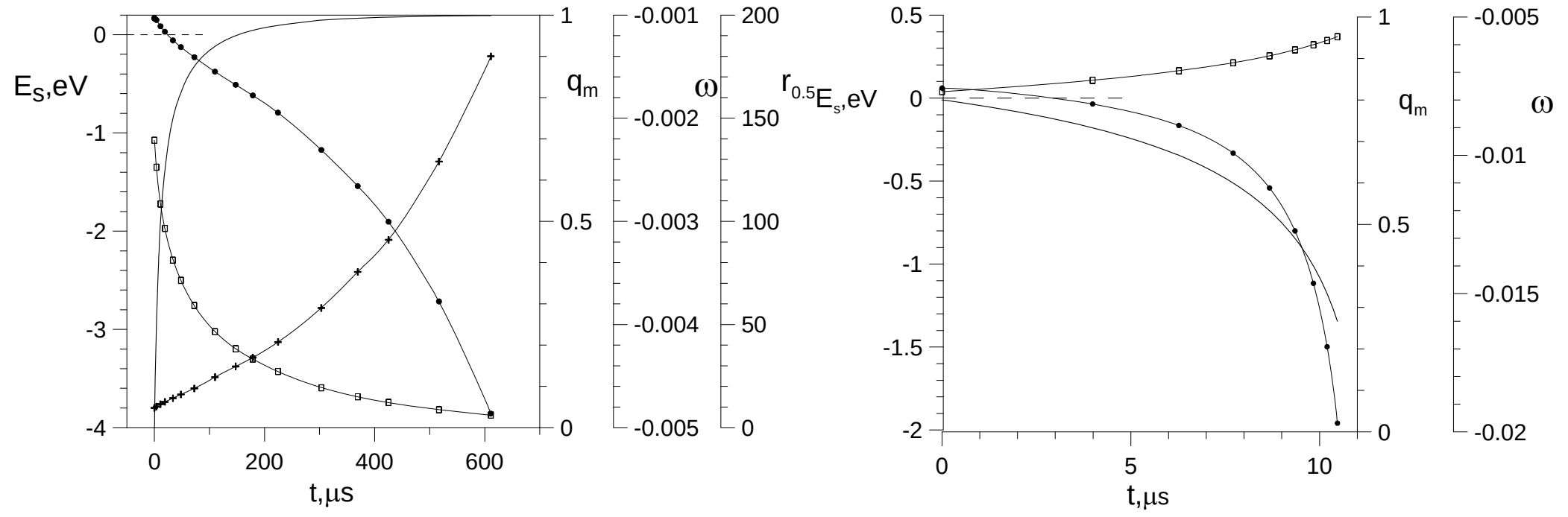

Fig.23 The time dependencies of energy, amplitude, radius and frequency (continuous line), of PBS if $h=1.001$ and the initial amplitude $q_{m,init}=0.696$ (see Fig.22).

Fig.24 The time dependencies of energy, amplitude, and frequency (continuous line) of PBS if $h=1.001$ and the initial amplitude $q_{m,init}=0.82$ (see Fig.21).

In Fig.24, the time dependencies of energy, amplitude, and precession frequency at $h=1.001$ , $q_{m,init}>\sqrt{0.5}$ are shown, when the PBS are transforming into the domains of the high-field phase.

## 5. The change of PBS during their movement

Now, we consider the influence of movement on soliton form. Using Eqs. (9), (11), and (12) for the $Q\omega=0$ case, we obtain the following expression:

$$\frac{\partial q}{\partial \tau}=\sqrt{k_1}\,2k\left(m_z\frac{\partial q}{\partial x}+l_z\frac{\partial p}{\partial x}\right)\cong 2k_1k(\omega+h)(3q^2-1)\frac{\partial q}{\partial x}. \tag{26}$$

Let us present the $q(x,y,z,\tau)$ function as $q(x,y,z,\tau)=q_S(x_S,y,z,\tau)$, where $x_S=(x-v_0\tau)$. In such a case, we have

$$-v_0\frac{\partial q_s}{\partial x_s}+\frac{\partial q_s}{\partial \tau}=-2k_1k(\omega+h)\frac{\partial q_s}{\partial x_s}+6k_1k(\omega+h)q^2\frac{\partial q_s}{\partial x_s}. \tag{27}$$

If we designate the value

$$v_0=2k_1(\omega+h)k. \tag{28}$$

as the velocity of movement of a soliton as a whole, then the expression $\frac{\partial q_s}{\partial \tau}=6k_1k(\omega+h)q^2\frac{\partial q_s}{\partial x_s}$ or, in spherical coordinates:

$$\frac{\partial q_s}{\partial \tau}=3v_0q^2\frac{\partial q_s}{\partial r_s}sin\,\theta\,cos\,\varphi \tag{29}$$

(here $r_S$ is the radial coordinate in the system of moving soliton) describes the deformation of a soliton because of its movement along the *x*-axis. For our solitons, the derivative $\frac{\partial q_S}{\partial r}<0$. Consequently, the PBS frontal side is decreasing, i.e. it becomes steeper, but the back side is increasing, i.e. it becomes more slope in the same extent. Thereby, "the centre of gravity" of soliton is displaced in a direction opposite to a direction of the main movement. Thus, the form of moving soliton differs from spherical, but if $Q=0$ its energy does not change.

However, the dissipation ($Q \neq 0$) results not only in change of frequency, but also in decrease of the velocity and, accordingly, of kinetic energy:

$$E_{kinet} = \frac{\pi M_0 B}{k_1(\omega + h)^2}\left(\frac{\alpha}{K_1}\right)^{1.5} v_0^2 \int q^2 r^2 dr \,. \qquad (30)$$

In turn, as a result of the velocity reduction, PBS asymptotically becomes more spherical in the form (more precisely, owing to anisotropy, ellipsoid of rotation).

## 6. Influence of the demagnetizing fields and a more correct equation for PBS

To estimate approximately the influence of demagnetizing field in the case of PBS, let's use the formula that concerns homogeneity magnetize sphere only:

$$H_{demag} = -\frac{4\pi}{3} M_z \,. \qquad (31)$$

Corresponding to Eqs. (14), (15), magnetization of PBS equals

$$M_s = 2M_0\sqrt{k_1}(\omega + h)q^2 \,. \qquad (32)$$

If to us the expression (32) for the magnetization in Eq.(31), we receive:

$$h_{demag} \cong -\frac{8\pi M_0}{3B}(\omega + h)q^2 \,. \qquad (33)$$

We can use this expression to estimate the value of demagnetizing field. With constants that were used in our examples, we receive $h_{demag} = -0.9\times 10^{-3} q^2$, i.e. the relative change of magnetic field $\Delta < 10^{-3}$ (in our cases $h \cong 1$). Such correction does not change qualitatively the characteristics of PBS and changes them quantitatively very little. Therefore, we can neglect demagnetizing fields.

**Formatted:** Indent: First line: 0 pt

It is possible to write down more correct equation for PBS, in comparison with Eq.(21), taking into account the additions proportional to $k_1$:

$$\left(1 + k_1(\omega + h)^2 q^2\right)\left(\frac{d^2 q}{dr^2} + \frac{2}{r}\frac{dq}{dr}\right) + \frac{q}{1 - q^2}\left[1 + 2k_1(\omega + h)^2 - k_1(\omega + h)^2 q^2\right]\left(\frac{dq}{dr}\right)^2 =$$
$$= q\left(1 - q^2\right)\left[\left(1 - \frac{k_2}{k_1}q^2\right)\left(1 + k_1(\omega + h)^2\left(1 - q^2\right)\right)\left(1 - k_1(\omega + h)^2 q^2\right) - (\omega + h)^2\right] \qquad (34)$$

However, using the value $k_1 = 1.43\times 10^{-4}$ corresponding to $Cr_2O_3$, the solutions of Eq.(34) differ less than by 1% from the solutions of Eq.(21).

## Conclusions

1. It is shown that dissipation of energy for precessing ball solitons is accompanied by the change of the precession frequency and by the deceleration of spatial movement.
2. Kinetics of the PBS is defined by the sign of initial precession frequency and the amplitude of the originated PBS. PBS transforms into the domains of the new phase if the initial frequency of precession is negative ($\omega_{init} < 0$) and the initial amplitude $q_{m,init} > \sqrt{0.5}$. In the paper, the whole process of such transformation of PBS has been analyzed.
3. If $h < 1$, the equilibrium PBS of two types are possible: precessing equilibrium solitons for which $\omega > 0$, present the first type; and non-precessing equilibrium solitons for which $\omega = 0$, the second type. In this paper, the process of evolution of PBS into these equilibrium PBS are analyzed. The PBS of the first type are most surprising, since in these cases the existence of stable precessing exitations at presence of dissipative member in equation of motion appears to be possible.
4. At $h > 1$ (or $h < \sqrt{1 - k_2/k_1}$ for reverse spin-flop transition), the so-called overcritical PBS are possible. It is supposed that such solitons can originate during the disintegration of the initial phase. For such solitons $\omega < 0$ (or $\omega > 0$ for reverse transition). If amplitude of initial PBS $q_{m,init} > \sqrt{0.5}$, such solitons transform into domains of the new phase. The relatively small-amplitude solitons, i.e. when the initial amplitude $q_{m,init} < \sqrt{0.5}$, the PBS decrease slowly to zero in amplitude but increase simultaneously in volume. The time dependencies of energy, amplitude, frequency, and radius of such solitons have been presented. It is possible to interpret such solitons originated during disintegration as "softening" of the initial phase.
5. At spatial movement of PBS, its form is deformed, but it does not change the size and the amplitude. The dissipation results not only in change of precession frequency, but also in reduction of the velocity of movement. As a result of the velocity reduction, the shape of PBS is approaching to spherical.

## Acknowledgment

Author is very grateful to Professor A. Zvezdin for useful discussions of the solitonic problem.

## References

[1] V.V.Nietz, Communication JINR No.P17-87-28 Dubna (1987) (in Russian).

[2] V.Yu.Bezzabotnov, V.V.Nietz, S.A.Oleynik, Communication JINR No.P17-95-87 Dubna (1995) (in Russian).

[3] V.V.Nietz, S.A.Oleynik, Communication JINR No.P17-95-88 Dubna (1995) (in Russian).

[4] V.V.Nietz, Non-linear periodic waves and solitons in an antiferromagnet with uni-axial anisotropy during a spin-flop transition, J. Moscow Phys. Soc. 9 (1) (1999) 63.

[5] V.V.Nietz, Neutron scattering by ball solitons, Euro-Asian Symposium "Trends in Magnetism", Ekaterinburg, Russia, February 27 - March 2. 2001; Physics of Metals and Metallography 92 Suppl.1 (2001) S243.

[6] V.V.Nietz, Neutron scattering by magnetic ball solitons, J. Magn. Magn. Mater. 266 (2003) 258.

[7] V.V.Nietz, A.A.Osipov, Ball solitons and kinetics of the first order phase transition, Crystallography Reports 53 (2) (2008) 266.

[8] V.V.Nietz, A.A.Osipov, Ball solitons in kinetics of the first order magnetic phase transition, J. Magn. Magn. Mater. 320 (2008) 1464.

[9] S.Foner, Phys.Rev. 130 (1) (1963) 183.